\documentclass[a4paper,
               ]{jacow}
               
\usepackage[USenglish]{babel}
\usepackage{upgreek}
\usepackage{amsmath}

\begin{document}

\title{Study of Collective Effects in the CERN FCC-ee Top-up Booster}

\author{D. Quartullo\thanks{Danilo.Quartullo@roma1.infn.it}, INFN-Rome1, Rome, Italy \\
M. Migliorati, Sapienza University and INFN-Rome1, Rome, Italy \\
M. Zobov, LNF-INFN, Frascati, Italy}
	
\maketitle

\begin{abstract}
   The CERN FCC-ee top-up booster synchrotron will accelerate electrons and positrons from an injection energy of 20 GeV up to an extraction energy between 45.6 GeV and 182.5 GeV depending on the operation mode. These accelerated beams will be used for the initial filling of the high-luminosity FCC-ee collider and for keeping the beam current constant over time using continuous top-up injection. Due to the high-intensities of the circulating beams, collective effects may represent a limitation in the top-up booster. In this work we present a first evaluation of the impedance model and the effects on beam dynamics.
   Methods to mitigate possible instabilities will be also discussed.
\end{abstract}

\section{INTRODUCTION}

The CERN $\text{e}^{+}\text{e}^{-}$ Future Circular Collider (FCC-ee) is a high-luminosity and high-precision electron-positron circular collider envisioned in a 100 km tunnel in the CERN-Geneva area \cite{Abada:2019zxq}. The FCC-ee will allow detailed studies of the heaviest known particles (Z, W, H bosons and the top quark) offering also great sensitivity to new particle physics.

The FCC-ee target luminosities of $10^{34}-10^{36}$~\si{\per\square\centi\meter\per\second} will lead to short beam lifetimes, due to beamstrahlung, radiative Bhabha scattering and Touschek effect \cite{Abada:2019zxq}. In order to sustain these short beam lifetimes, a full-energy booster, installed in the collider tunnel, will provide continuous top-up injection, in addition to initially filling the FCC-ee.  

The booster will be built in the same tunnel used for the collider and the circumference lengths of the two machines will be the same, almost 100 km. The booster will accelerate batches of electrons and positrons from an injection energy of 20 GeV up to an extraction energy of 45.6 GeV, 80~GeV, 120 GeV, 182.5 GeV respectively for Z, W, H and top quark productions \cite{Abada:2019zxq}. This design injection energy, which corresponds to a magnetic field $B=6$ mT, could change in the future, depending on the quality and reproducibility of the magnetic field in the dipole magnets.

In order to not affect the collider luminosity and to diminish the background generated by lost particles, the booster is expected to provide at extraction an equilibrium transverse emittance similar to the one in the collider. Since the FCC-ee lattice will be optimized for two optics, one with 60° phase advance for the Z and W experiments, the other with 90° phase advance for the H and top quark productions, the optics in the booster will change depending on the phase advance in the collider.

The synchrotron radiation (SR) transverse damping time at booster injection-energy will be longer than 10 s, leading to incompatibility with the booster cycle \cite{Harer:2649814}. In addition the relatively small horizontal normalized equilibrium emittance of 12 pm rad will cause emittance blow-up along the cycle due to intra-beam scattering. In order to solve these issues, 16 wigglers should be installed in the booster, leading to a damping time of 0.1 s and an emittance of 240 pm rad and 180 pm rad for the 60° and 90° optics respectively.

The relatively high nominal intensity of $N_\text{b}=3.4\times 10^{10}$ particles per bunch (ppb) could lead to collective effects able to severely limit the booster operation. In particular, the resistive wall effect due to the foreseen beam-pipe in stainless steel and with radius $r_\text{c}=25$ mm could cause strong instabilities in both longitudinal and transverse planes.

A first evaluation of the importance of collective effects in the booster at injection energy was reported in Ref.\cite{mauro}, where it was shown that, in the absence of wigglers, an intensity threshold of $0.1\times 10^{10}$ ppb, significantly lower than the nominal intensity, was defined by the microwave instability (MI) caused by the resistive wall. In the transverse plane, the intensity threshold due to transverse mode-coupling instability (TMCI) was only $0.6\times 10^{10}$ ppb. Moreover, analytical estimations of the resistive wall transverse coupled-bunch instability (TCBI) found a rise time of just few revolution turns which requires new feedback schemes \cite{Drago:2017dwa}.

This work aims at finding possible cures to the impedance-induced instabilities in the booster, focusing on the beam dynamics at injection energy and assuming an optics with 60° phase advance and no wigglers installed in the machine. 

The next section highlights some significant machine and beam parameters considered in the present study. Then careful estimations of the resistive-wall impedance are given, together with a possible way to lower it. The next step is to study the MI in the longitudinal plane through macro-particle simulations, providing a possible cure to increase the intensity threshold. Choosing a proper combination of parameters which allows having stable beams at the nominal intensity, the study then shifts to the transverse plane, where a semi-analytical Vlasov solver is used to find the TMCI intensity threshold. Finally, estimations of the TCBI rise-time are provided using analytical formulae.

\section{Machine and beam parameters considered in the study}

Table \ref{tab:sim} shows some significant machine and beam parameters used in the study of collective effects in the booster. 

\begin{table}[!h]
	\centering
	\caption{Main machine and beam parameters relevant for the booster studies. Most of these quantities can be found in Ref.\cite{Abada:2019zxq}.}
	\begin{tabular}{lcc}
		\toprule
		\textbf{Parameter} & \textbf{Value} \\
		\midrule
		Machine circumference ($C_\text{r}$)         & \SI{97.756}{km} \\ 
		Beam energy at injection ($E_0$)         & \SI{20}{GeV} \\ 
		Beam rev. frequency at injection ($f_0$)         & \SI{3.06}{kHz} \\
		Bunch population ($N_\text{b}^{\text{nom}}$)         & $3.4\times 10^{10}$ ppb \\
		Number of bunches per beam ($M_\text{b}$)         & 16640 \\ 
		SR 1 $\sigma$ rel. energy spread ($\sigma_{dE0,\text{r}}$)         & $0.166\times 10^{-3}$ \\ 
		SR energy loss per turn ($U_0$)         & $1.33$ MeV \\ 
		SR damping time ($\tau_z$)         & 15013 turns \\ 
		SR 1 $\sigma$ bunch length ($\sigma_{z0}$)         & 1.26 mm \\ 
		RF frequency ($f_{\text{rf}}$)  & 400 MHz \\ 
		Harmonic number ($h$) & 130432 \\ 
		RF voltage ($V_{\text{rf}}$) & 60 MV \\ 
		Arc phase advance ($\phi_{\text{a}}$) & 60° \\ 
		Momentum compaction factor ($\alpha_{\text{c}}$) & $1.48\times 10^{-5}$ \\
		Synchrotron tune ($Q_{s}$) & 0.0304 \\
		Betatron tunes ($Q_{x,y}^{\text{nom}}$) & 269.139 \\
		\bottomrule
	\end{tabular}
	\label{tab:sim}
\end{table}

As discussed later, the beam dynamics at injection energy is the most critical as concerns intensity effects. 

The considered $\phi_\text{a}=$ 60° directly determines $\alpha_c$ and influences $Q_s$ and $\sigma_{z0}$. As shown later, the MI intensity threshold is proportional to $\alpha_\text{c}$ and $\sigma_{z0}$ while the TMCI intensity threshold is proportional to $Q_s$. A phase advance of 90° would lead to lower values for $\alpha_\text{c}$, $Q_s$, $\sigma_{z0}$ and therefore to more critical scenarios.

The main three SR parameters $\sigma_{z0}$, $U_0$ and $\tau_z$ can be directly computed assuming an average dipole-magnet bending radius $\rho_\text{b}=10.6$ km in the absence of wigglers. 

The booster main RF system consists of 400 MHz superconducting cavities \cite{Abada:2019zxq}. In the present work a voltage $V_{\text{rf}}=60$ MV at injection energy is assumed. This value is significantly higher than $U_0$ so that the bunch practically sees the linear part of the RF voltage at each revolution turn. In addition, the chosen $V_{\text{rf}}$ is expected to be substantially lower than the total available voltage, so that $V_{\text{rf}}$ can be increased during acceleration according to the cycle needs.

The number of bunches simultaneously accelerated in the booster will depend on the collider experiment. Table~1 indicates the largest planned $M_\text{b}$, which relates to the Z-boson experiment \cite{Abada:2019zxq}. 

Concerning the transverse plane, the horizontal and vertical tunes are assumed to be equal to the horizontal tune foreseen in the FCC-ee for the Z-boson experiment \cite{Abada:2019zxq}.

\section{Estimation of the resistive wall impedance}

The resistive wall impedance is the dominant component of the booster impedance model and it is the only impedance contribution considered in this paper.

\begin{figure}[!h]
	\includegraphics[width=\columnwidth]{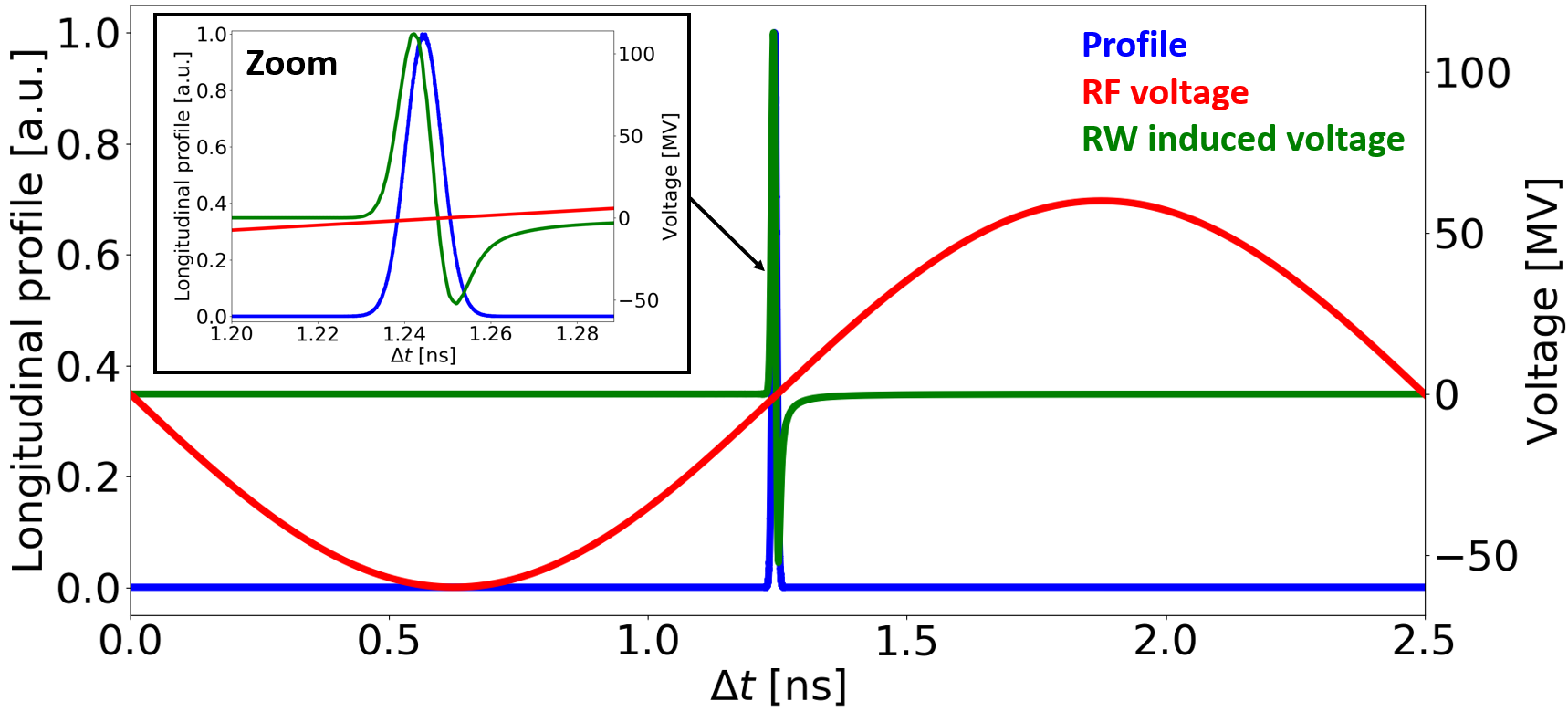}
	\caption{Longitudinal profile (blue), RF voltage (red) and resistive wall induced voltage (green) as a function of time in the FCC-ee booster considering the parameters in Table 1 and a stainless steel beam pipe with radius $r_\text{c}=25$~mm.}
	\label{fig:RWind}
\end{figure}

The baseline for costs minimization is to have a circular beam-pipe in stainless steel (resistivity $\rho_\text{r}=7\times 10^{-7} \Omega$m) with radius $r_\text{c}=25$ mm. However the corresponding resistive wall impedance would lead to a peak induced voltage even higher than the peak RF voltage for a bunch with the nominal intensity, see Fig.\ref{fig:RWind}.

\begin{figure}[!h]
	\includegraphics[width=\columnwidth]{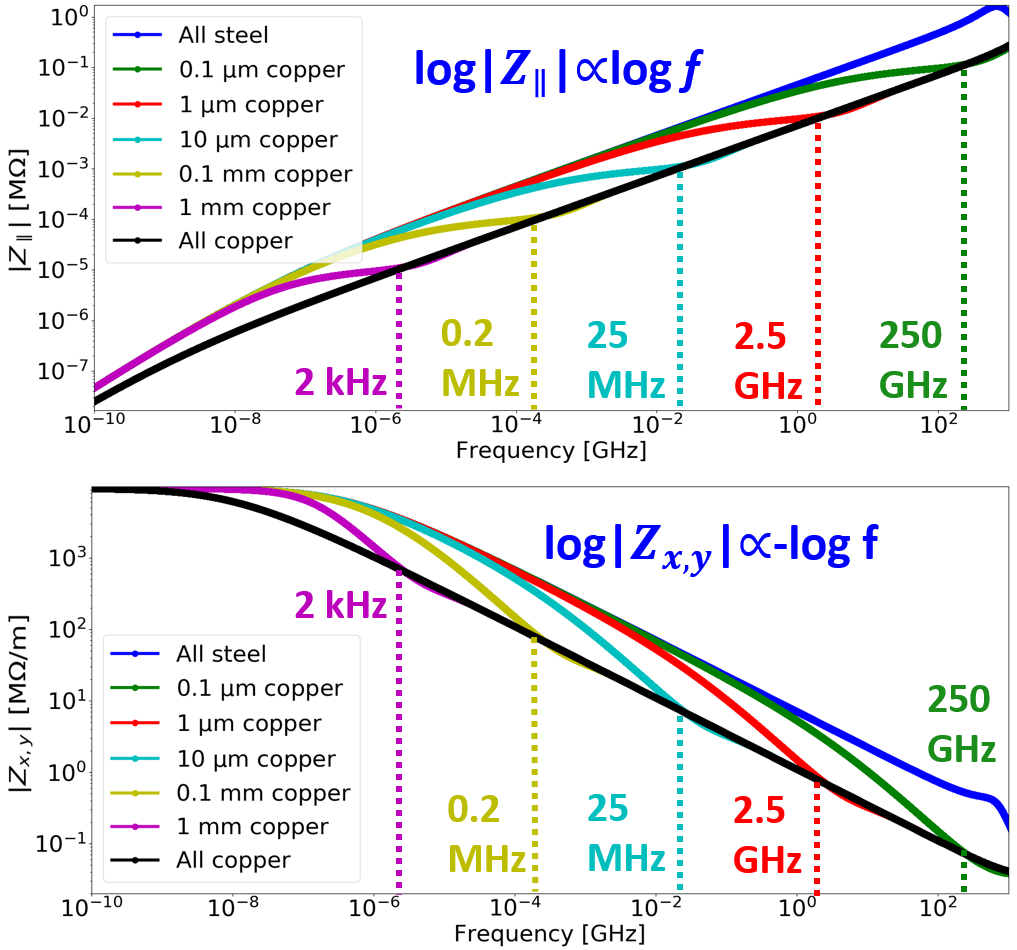}
	\caption{Longitudinal (top) and transverse dipolar (bottom) resistive wall impedance as a function of frequency in the FCC-ee booster assuming a beam-pipe in stainless steel with radius $r_\text{c}=25$ mm and different values for the thickness of the internal copper layer. The five vertical dashed lines mark the frequencies $f_{\delta_\text{l}}$ above which the corresponding impedances converge to the single-layer copper impedance. The dependences of the single-layer impedances on the frequency are also reported in both plots. All the impedance curves have been obtained with the IW2D code.}
	\label{fig:longRW}
\end{figure}

In order to significantly reduce this impedance, the possibility of applying a cooper coating to the beam-pipe was investigated. Specifically, the CERN IW2D code \cite{iw2d} was used to evaluate the longitudinal and transverse resistive wall impedances of a two-layer vacuum chamber, being the external layer in stainless steel and the internal layer in copper ($\rho_\text{r}=1.7\times 10^{-8} \Omega$m) with variable thickness $\delta_\text{l}$, see Fig.\ref{fig:longRW}. 

For single-layer beam-pipes, the dependences of the longitudinal and transverse-dipolar resistive wall impedances on the frequency are \cite{Henry:1991xi}
\begin{equation}\label{conto}
|Z_\parallel(f)|\propto \frac{1}{r_c}\sqrt{\rho_r f}, \quad |Z_{x,y}(f)|\propto \frac{1}{r_c^3}\sqrt{\frac{\rho_r}{f}}
\end{equation}
in a certain asymptotic range of frequencies. These dependencies are highlighted in Fig.\ref{fig:longRW}, where all the axes are in logarithmic scale. 

The plots in Fig.\ref{fig:longRW} also show that, for a given $\delta_\text{l}$, both longitudinal and transverse impedances converge to the single-layer stainless-steel and copper impedances respectively for low and high frequencies. In particular, for a certain $\delta_\text{l}$, the frequencies above which the longitudinal and transverse impedances converge to the single-layer copper impedance are essentially the same and will be denoted by $f_{\delta_\text{l}}$ below. 

In the next Section the MI thresholds will be evaluated for the different impedances shown in Fig.\ref{fig:longRW} (top). As reported in Ref.\cite{Shaposhnikova:381423}, MI can occur when the wavelength of the wakefield is much shorter than the bunch length, i.e. $f_\text{c} \tau \gg 1$, where $f_\text{c}$ is the frequency of the impedance which drives MI and $\tau$ is the full bunch length. Therefore it is expected that the impedance related to a given $\delta_\text{l}$ and the impedance of the single-layer copper beam-pipe will lead to the same MI intensity threshold if these two impedances coincide for frequencies larger than $1/\tau$, i.e. when $f_{\delta_\text{l}}<1/\tau$. 

Finally it should be noted that, being the booster a fast-cycling machine, eddy currents in presence of a copper layer could be an issue during acceleration and their effects should be separately investigated.

\section{Cures for increasing the microwave-instability intensity threshold}

Single-bunch macroparticle longitudinal beam dynamics simulations were performed with the CERN BLonD code \cite{blond} in order to evaluate the MI intensity threshold $N_{b,\text{th}}^{\text{MI}}$ taking into account the parameters of Table 1 and the resistive-wall impedance shown in Fig.\ref{fig:longRW} (top) with a variable $\delta_\text{l}$.

Simulation tracking lasted $10^{6}$ revolution turns, which is more than 6 times the SR longitudinal damping time, in such a way to reach a SR equilibrium at the end of simulations. It should be noted that $10^{6}$ turns corresponds to 32.6 s, which is comparable to the flat-bottom duration of 51.1 s in the booster for the Z-boson experiment \cite{Abada:2019zxq}.

\begin{figure}[!h]
	\includegraphics[width=\columnwidth]{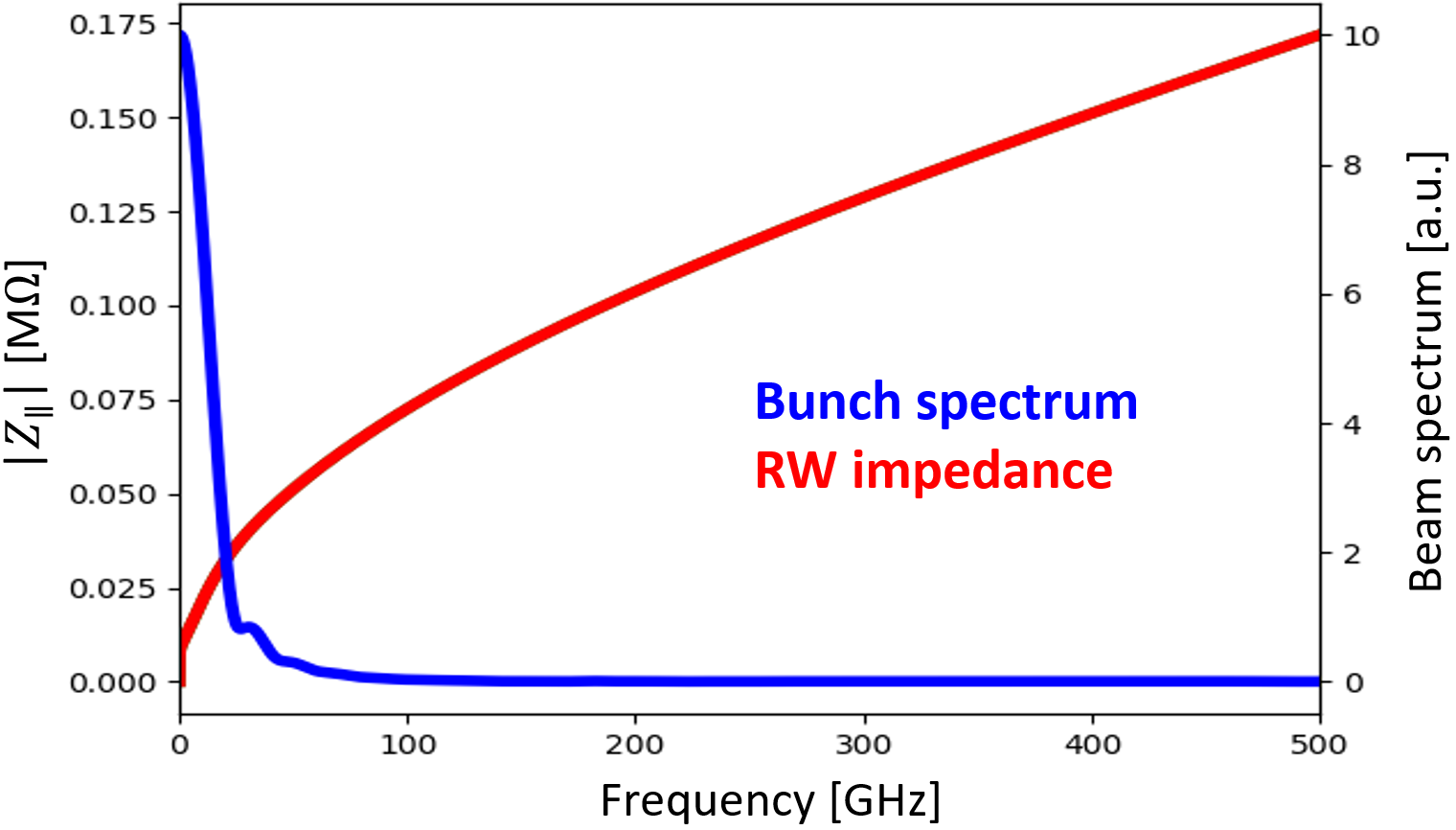}
	\caption{Bunch spectrum and resistive wall impedance used to compute the induced voltage in BLonD simulations. Case of $\delta_\text{l}=1 \upmu$m, $\sigma_{z}=4$ mm at SR equilibrium.}
	\label{fig:fourier}
\end{figure}

The resistive-wall induced voltage was computed in frequency domain, multiplying the bunch spectrum by the impedance and performing an inverse Fourier transform. Due to the relatively short bunches and in order to have an acceptable resolution in the longitudinal profile binning (at least 50 slices for $4 \sigma_{z}$), the maximum frequency considered in computations was 500 GHz. This obliged to use a relatively large number of macroparticles per bunch (more than $10^{7}$) in order to counteract the numerical noise obtained when multiplying impedance and spectrum, which are respectively increasing and decreasing functions of frequency (Fig.\ref{fig:fourier}).

\begin{figure}[!h]
	\includegraphics[width=\columnwidth]{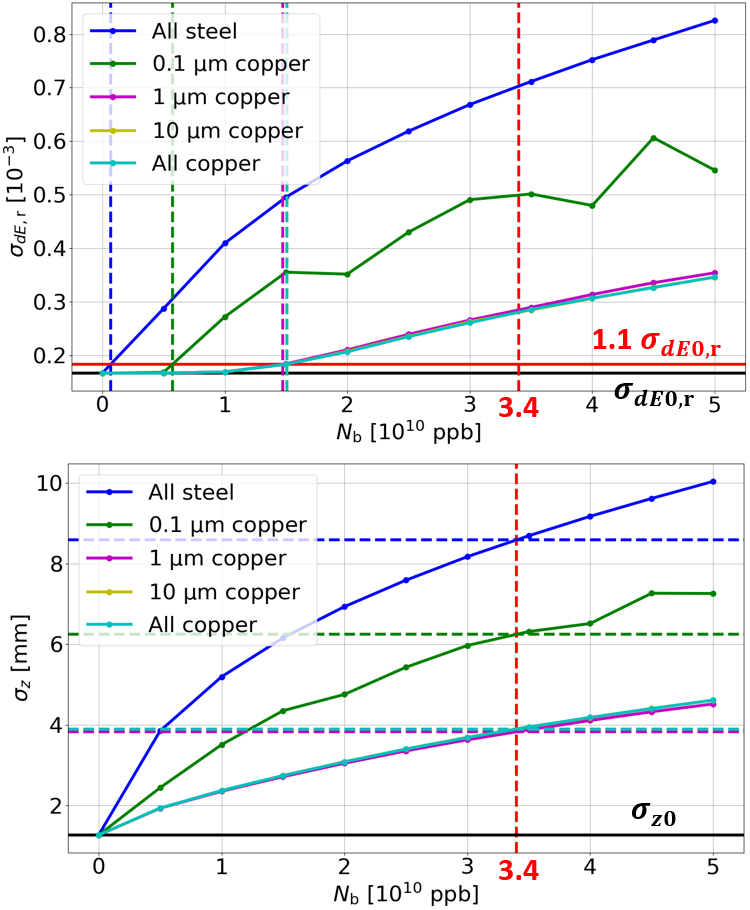}
	\caption{Equilibrium rms relative energy spread (top) and bunch length (bottom) as a function of bunch population obtained with BLonD simulations. The used parameters are reported in Table 1 and the resistive wall impedance shown in Fig.\ref{fig:longRW} (top) has been included in simulations varying $\delta_\text{l}$. Top: the horizontal lines mark $\sigma_{dE0,\text{r}}$ and its increase by 10\%. The vertical lines mark $N_{b,\text{th}}^{\text{MI}}$ for the corresponding curves. Bottom: the horizontal lines mark $\sigma_{z0}$ and the values of $\sigma_{z}$ when $N_\text{b}=3.4\times 10^{10}$ ppb and $\delta_\text{l}$ varies. In both images the yellow and cyan curves are overlapped.}
	\label{fig:mi}
\end{figure}

Figure \ref{fig:mi} shows the simulation results, specifically the  equilibrium $\sigma_{dE,\text{r}}$ and $\sigma_{z}$ (averaged over the last 10000 revolution turns) as a function of $N_\text{b}$ and varying $\delta_\text{l}$.

When observed in simulation, MI led to a $\sigma_{dE,\text{r}}$ increase relative to $\sigma_{dE0,\text{r}}$. More specifically, up to a certain intensity threshold $N_{\text{b},\text{th}}^{\text{MI}}$, $\sigma_{dE,\text{r}}\approx \sigma_{dE0,\text{r}}$, while $\sigma_{dE,\text{r}}$ becomes an increasing function of $N_\text{b}$ when $N_\text{b}>N_{\text{b},\text{th}}^{\text{MI}}$. In Fig.\ref{fig:mi} (top), in order to assess an unambiguous way to determine $N_{\text{b},\text{th}}^{\text{MI}}$, the MI intensity threshold is chosen so that, when  $N_\text{b}=N_{\text{b},\text{th}}^{\text{MI}}$, then $\sigma_{dE,\text{r}}=1.1 \sigma_{dE0,\text{r}}$. 

Figure \ref{fig:mi} (top) shows that, even with a beam-pipe entirely made of copper, the threshold $N_{\text{b},\text{th}}^{\text{MI}}$ is only $1.5\times 10^{10}$ ppb, significantly lower than the nominal bunch intensity. This largest value for $N_{\text{b},\text{th}}^{\text{MI}}$ can be also obtained with good approximation when $\delta_\text{l}=1 \upmu$m. 

Concerning the equilibrium bunch length, Fig.\ref{fig:mi} (bottom) shows that $\sigma_{z}$ is an increasing function of $N_\text{b}$ and the bunch lengthening relative to $\sigma_{z0}$ is higher when the resistive-wall impedance is larger (smaller $\delta_\text{l}$). This bunch lengthening occurs even when $N_\text{b}<N_{\text{b},\text{th}}^{\text{MI}}$ and no clear changes in curve behaviour are visible in correspondence of $N_\text{b}=N_{\text{b},\text{th}}^{\text{MI}}$.

With a vacuum-chamber entirely in copper, the equilibrium full bunch-length is $\tau\approx 4\sigma_z = 36$ ps when $N_\text{b}=1.5\times 10^{10}$~ppb (Fig.\ref{fig:mi}, bottom). Therefore, following the reasoning of the previous Section, a copper coating should lead to the highest possible MI intensity-threshold ($1.5\times 10^{10}$~ppb) when $f_{\delta_\text{l}}<1/\tau\approx 30$ GHz. This is in perfect agreement with the $f_{\delta_\text{l}}$ values reported in Fig.\ref{fig:longRW} (top), since $f_{\delta_\text{l}}$ is larger than 30 GHz when $\delta_\text{l}=0.1 \upmu$m and lower than 30 GHz when $\delta_\text{l}=1 \upmu$m.   

Since the reduction of the longitudinal resistive-wall impedance by adding a copper layer to the beam-pipe did not avoid MI for the nominal bunch intensity, a second cure for instability has been studied. 

For the Boussard criterion \cite{Boussard:872559}, $N_{\text{b},\text{th}}^{\text{MI}}$ scales as
\begin{equation}\label{boussard}
N_{\text{b},\text{th}}^{\text{MI}}\propto \frac{\alpha_c E_0 \sigma_{dE0,\text{r}}^2 \sigma_{z0}}{|Z_\parallel|/n},
\end{equation}
where $n=f/f_0$. As expected, this expression shows that the MI intensity threshold increases when the longitudinal resistive wall impedance is reduced. Equation (\ref{boussard}) clarifies also the observation done above concerning the major strength of MI at booster injection-energy with the 90° phase-advance optics (lower $\alpha_c$). Even more importantly, Eq.(\ref{boussard}) shows that $N_{\text{b},\text{th}}^{\text{MI}}$ depends quadratically on $\sigma_{dE0,\text{r}}$ and linearly on $\sigma_{z0}$.

One way to increase $\sigma_{dE0,\text{r}}$ consists in installing wigglers in the booster with a consequent increase in $U_0$. Indeed, the two scaling relations \cite{4327284}
\begin{equation}\label{scaling}
\tau_z = \frac{1}{U_0(1+C_1 U_0)}, \quad \sigma_{dE0,\text{r}}\propto \sqrt{\frac{U_0}{1+C_2 U_0}}, 
\end{equation}
where $C_1$ and $C_2$ are positive quantities not depending on $U_0$, show that a larger $U_0$ leads to lower $\tau_z$ and higher $\sigma_{dE0,\text{r}}$.

Therefore the BLonD simulations described above were repeated varying $U_0$ and considering $\delta_\text{l}=1 \upmu$m, which is a good compromise between an increase in $N_{\text{b},\text{th}}^{\text{MI}}$ (Fig.\ref{fig:mi}, top) and a decrease in production costs and potential eddy-current issues related to the copper-layer thickness.

\begin{figure}[!h]
	\includegraphics[width=\columnwidth]{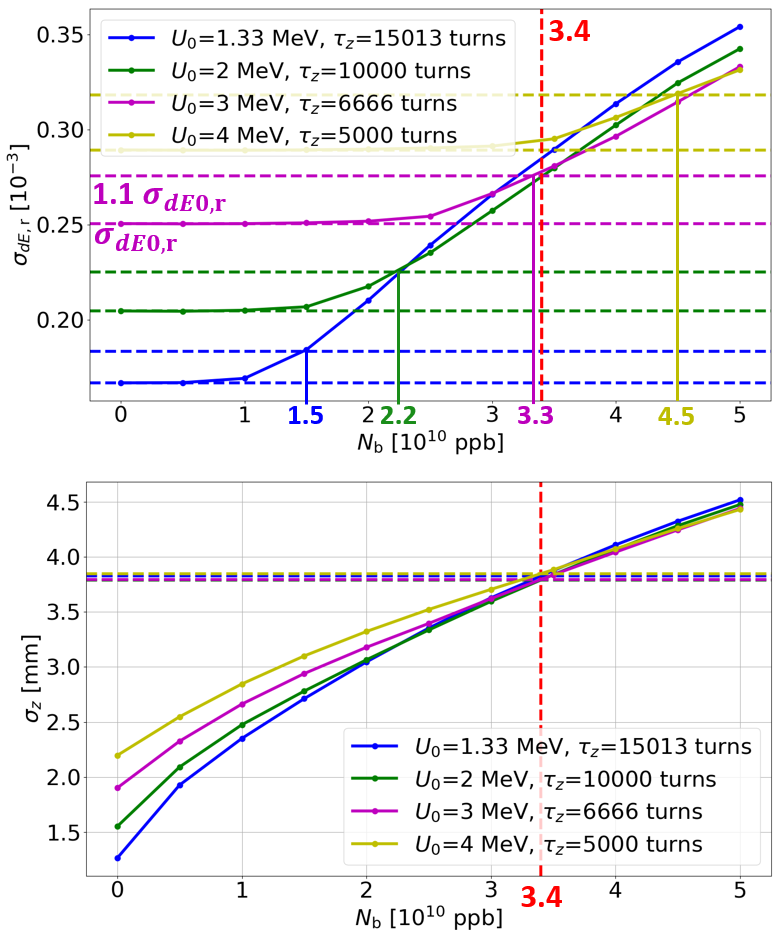}
	\caption{Equilibrium rms relative energy spread (top) and bunch length (bottom) as a function of bunch population obtained with BLonD simulations. The used parameters are reported in Table 1, except for the SR quantities which vary following Eq.(\ref{scaling}). The resistive wall impedance with $\delta_\text{l}=1 \upmu$m (Fig.\ref{fig:longRW}, top) has been included in simulations. Top: for each curve, the corresponding horizontal lines mark $\sigma_{dE0,\text{r}}$ and its increase by 10\%, while the vertical line marks $N_{\text{b},\text{th}}^{\text{MI}}$. Bottom: the horizontal lines mark the values of $\sigma_{z}$ when $N_\text{b}=3.4\times 10^{10}$ ppb and $U_0$ varies. }
	\label{fig:mi2}
\end{figure}

Figure \ref{fig:mi2} (top) shows the new intensity thresholds. As foreseen by Eqs.(\ref{boussard}) and \ref{scaling}, the plot shows that $N_{\text{b},\text{th}}^{\text{MI}}$ increases with $U_0$. In particular, $U_0$ should be at least 4 MeV in order not to have MI for the nominal bunch intensity. 

Regarding the bunch lengths, Fig.\ref{fig:mi2} (bottom) shows that $\sigma_{z0}$ increases as a function of $U_0$, helping in increasing $N_{\text{b},\text{th}}^{\text{MI}}$ as Eq.(\ref{boussard}) suggests. The plot also indicates that the increase of $\sigma_{z}$ with $U_0$ is less and less significant as $N_\text{b}$ approaches the nominal bunch intensity.

This second additional cure for MI, i.e. installing wigglers in the booster to increase $U_0$, is in full agreement with the current machine design-plan. Indeed, as already mentioned in the Introduction, considerations entirely concerning the transverse plane would lead to the installation of 16 wigglers able to providing $U_0=126$ MeV \cite{Abada:2019zxq, Harer:2649814}. Simulations with such a value for $U_0$ will require a much larger $V_{\text{rf}}$, moreover there will be likely no need to add a copper layer to the beam pipe due to the larger $\sigma_{dE0,\text{r}}$ in Eq.(\ref{boussard}). Further studies are needed to cover this scenario.

\section{Transverse mode-coupling instability intensity threshold}

The previous Section showed that a bunch with nominal intensity does not suffer MI when $U_0=4$ MeV and $\delta_\text{l}= 1\upmu$m. Taking into account these two conditions, beam dynamics studies in the transverse plane were needed to verify that the bunch will not be unstable due to TMCI. 

The intensity threshold for TMCI scales as \cite{Sacherer:322645}
\begin{equation}\label{sach}
N_{\text{b},\text{th}}^{\text{TMCI}}\propto \frac{Q_{x,y} Q_s E_0 \sigma_z}{\operatorname{Im}(Z_{x,y})},
\end{equation}
where $\operatorname{Im}(Z_{x,y})$ is the imaginary part of the transverse resistive-wall impedance. 

The quantity $\operatorname{Im}(Z_{x,y})$ was decreased adding a copper-layer to the beam pipe in order to cope with MI. Equation (\ref{sach}) shows a linear dependence of $N_{\text{b},\text{th}}^{\text{TMCI}}$ on the equilibrium $\sigma_z$ which, for a certain $U_0$, increases with $N_\text{b}$ due to SR and MI (see bottom plots in Figs.\ref{fig:mi} and \ref{fig:mi2}). This bunch-lengthening in the longitudinal plane helps increasing $N_{\text{b},\text{th}}^{\text{TMCI}}$. Note also that, at least for intensities close or above the nominal value, increasing $U_0$ would not lengthen the bunch (Fig.\ref{fig:mi2}, bottom) and therefore would not help to counteract TMCI.

The CERN DELPHI code \cite{delphi}, which is an analytical Vlasov solver for impedance-driven modes, was used to evaluate $N_{\text{b},\text{th}}^{\text{TMCI}}$ without taking into account the radiation damping. The values for $\sigma_z$ needed in DELPHI have been taken from the BLonD simulation results described above (yellow curve in the bottom plot of Fig.\ref{fig:mi2}) .

\begin{figure}[!h]
	\includegraphics[width=\columnwidth]{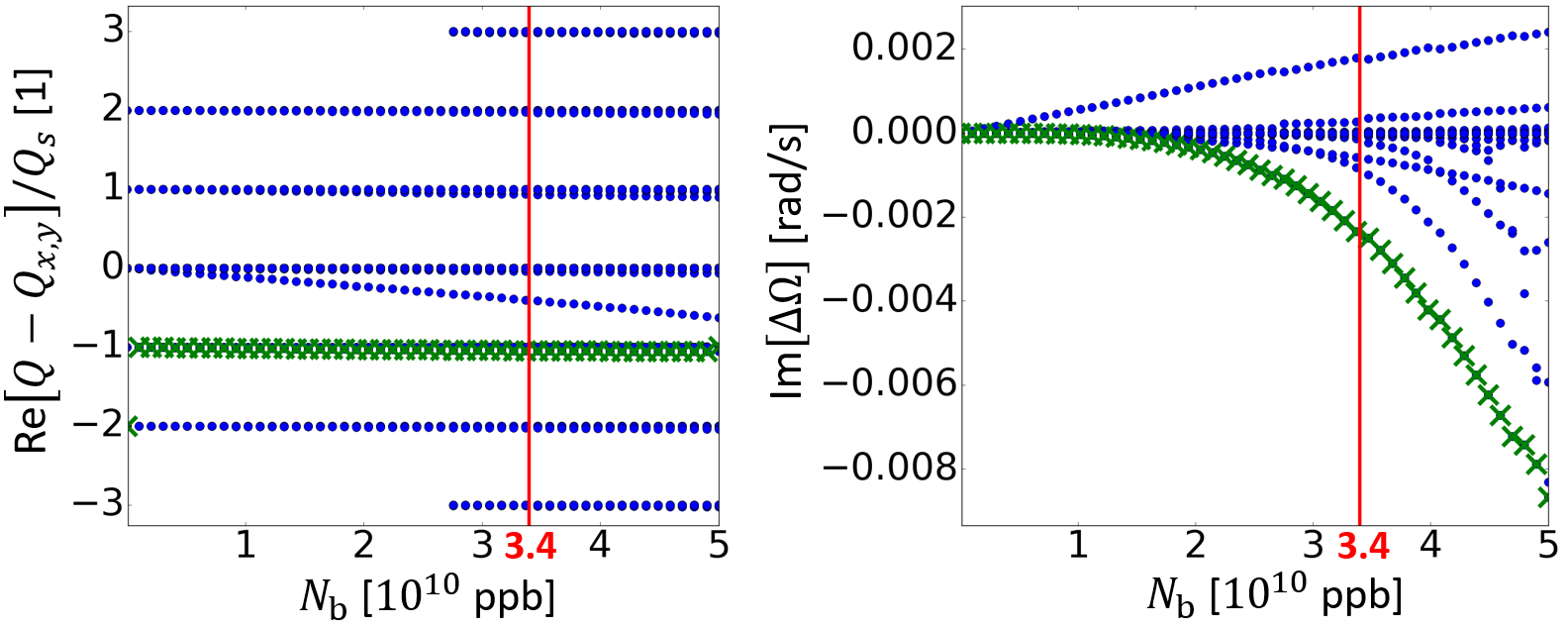}
	\caption{Real (left) and imaginary (right) parts of the tune shift of the first coherent oscillation modes as a function of the bunch population obtained with DELPHI. In both plots the green cross associated to a given $N_\text{b}$ marks the unstable mode with largest growth rate. The parameters needed in simulation are taken from Table 1, except for the equilibrium $\sigma_z$ which depends on $N_\text{b}$ according to the yellow curve in Fig.\ref{fig:mi2} (bottom, $U_0=4$ MeV). The value $\delta_\text{l}= 1\upmu$m is assumed for the dipolar resistive-wall impedance.}
	\label{fig:delphi}
\end{figure}

Figure \ref{fig:delphi} shows the real and imaginary parts of the complex tune shift of the first coherent oscillation modes obtained with DELPHI. 

No mode coupling occurs up to $N_\text{b}=5\times 10^{10}$ ppb (Fig.\ref{fig:delphi}, left). In addition the rise times of the unstable modes (Fig.\ref{fig:delphi}, right) are larger than 125 s, and these values are long compared to the SR transverse damping-time (3.26 s) and the flat-bottom durations in the booster, which vary from 1.6~s to 51.1 s according to the collider experiment \cite{Abada:2019zxq}.

\begin{figure}[!h]
	\includegraphics[width=\columnwidth]{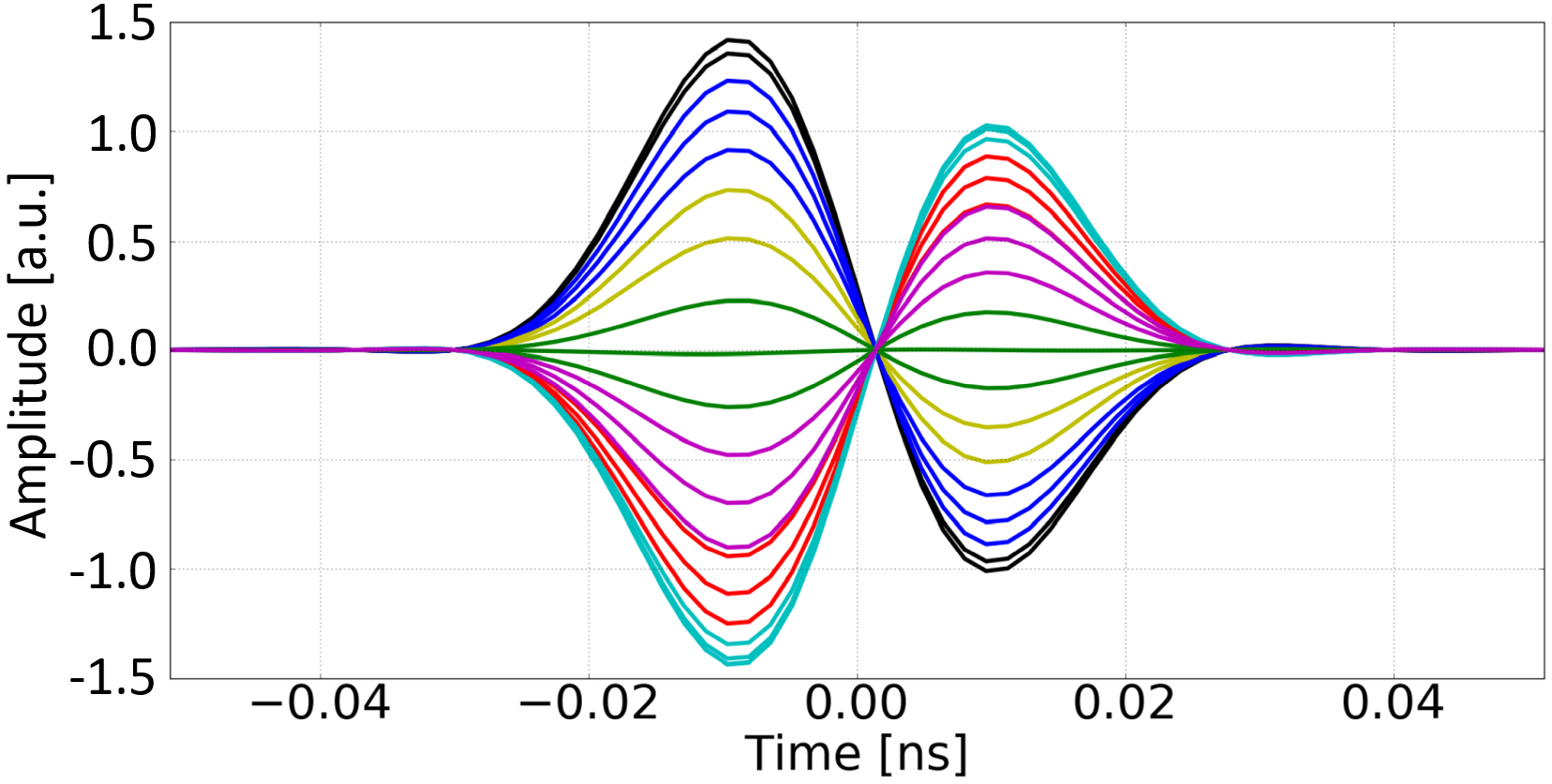}
	\caption{Twenty consecutive Head-Tail signals as a function of the longitudinal coordinate (0 ns corresponds to the bunch centre). These signals were obtained with DELPHI and refer to the simulations results shown in Fig.\ref{fig:delphi} when $N_\text{b}=N_\text{b}^{\text{nom}}$.}
	\label{fig:ht}
\end{figure}

Therefore $N_{\text{b},\text{th}}^{\text{TMCI}}>5\times 10^{10}$ ppb and, in particular, no TMCI is observed for the nominal bunch intensity. Figure~\ref{fig:ht} shows 20 consecutive Head-Tail signals obtained with the DELPHI code when $N_\text{b}=N_\text{b}^{\text{nom}}$. Notice that mode -1 largely prevails while mode 0 only creates a relatively small asymmetry between the amplitudes of the two signal halves.

\section{Resistive-wall transverse coupled-bunch instability}

The longitudinal resistive-wall wakefield decays along a distance much shorter than the booster-bucket length. On the contrary, the transverse resistive-wall wakefield is long-range and can lead to TCBI.

In the following, analytical estimations of the TCBI growth-rate are provided. As for the previous Section, all the needed parameters are taken from Table 1 and $\delta_\text{l}= 1\upmu$m as concerns the resistive-wall impedance.

Assuming $M_\text{b}$ equally-spaced bunches in the ring, the motion of the entire beam can be considered as the sum of $M_\text{b}$ coherent coupled-bunch modes. The transverse growth-rate $\alpha_{\mu}$ for the $\mu$-th coupled-bunch mode, where $\mu$ is an integer between 0 and $M_\text{b}-1$, can be easily computed taking into account only the most prominent radial mode in the azimuthal $m=0$ and assuming Gaussian bunches. The expression for $\alpha_{\mu}$ is \cite{doi:10.1142/5835}
\begin{equation}\label{alpha}
\alpha_{\mu} = - \frac{c e M_\text{b} N_\text{b} f_0}{4 \pi E_0 Q_{x,y}} \sum_{q=-\infty}^{\infty} \operatorname{Re}\left[Z_{x,y}\left(f_{\mu,q}\right)\right],
\end{equation}
where $E_0$ is in eV units and $f_{\mu,q}=f_0(q M_\text{b}+\mu+Q_{x,y})$.

\begin{figure}[!h]
	\includegraphics[width=\columnwidth]{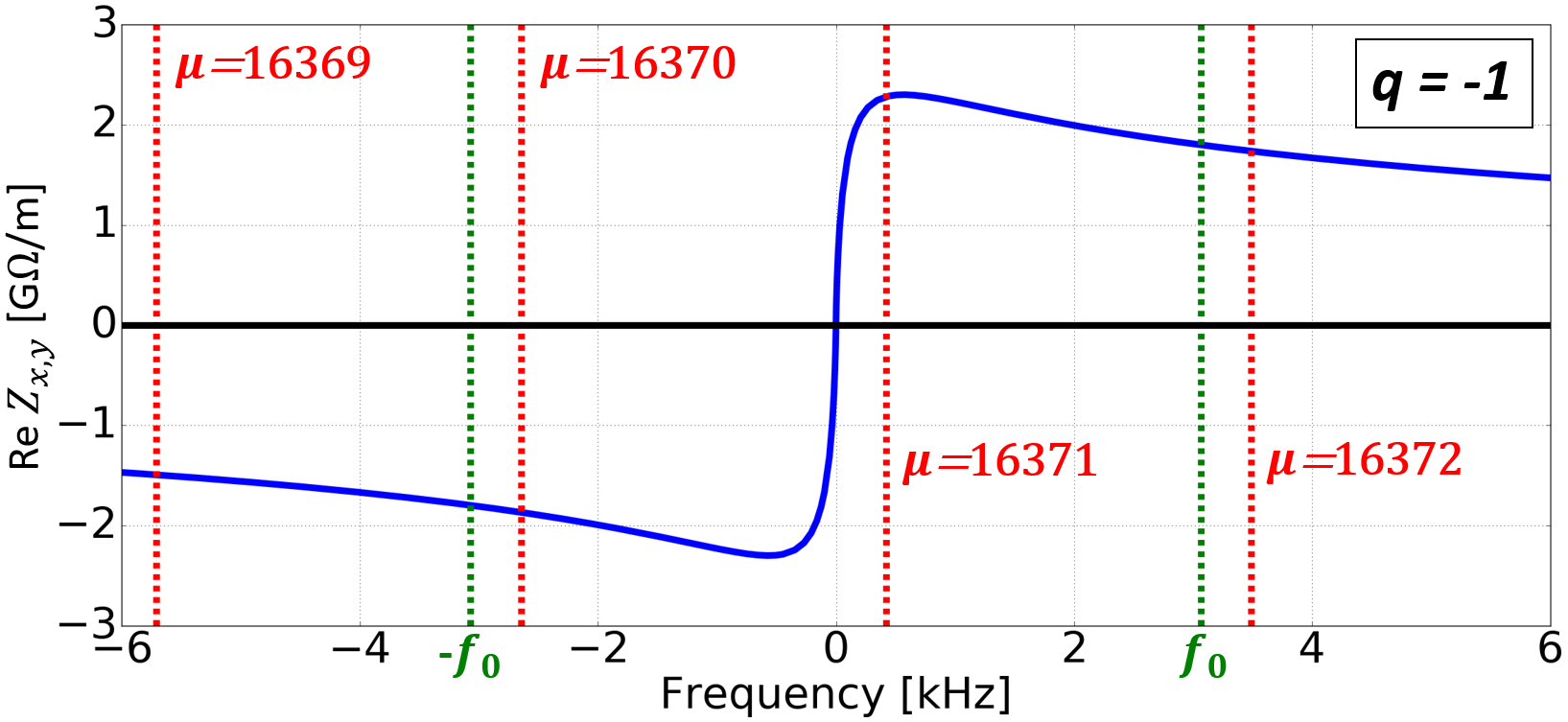}
	\caption{Resistive-wall transverse dipolar impedance (blue) as a function of frequency assuming $\delta_\text{l}= 1\upmu$m. The two green  lines mark plus or minus $f_0$. The four red lines mark $f_{\mu,q}$, where $q=-1$, $Q_{x,y}=Q_{x,y}^{\text{nom}}$ and $\mu$ varies as shown.}
	\label{fig:tcbi}
\end{figure}

Considering the value of $Q_{x,y}^{\text{nom}}$ and the shape of the transverse resistive-wall impedance near plus or minus $f_0$, it can be seen that the most unstable coupled-mode $\bar{\mu}$ satisfies the condition $(q M_\text{b}+\bar{\mu}+Q_{x,y}^{\text{nom}})\in [-1,0]$ for a certain $q$ (Fig.\ref{fig:tcbi}). This condition is satisfied for $\bar{\mu}=16370$ and $q=-1$. It is worth noting that for the most stable and unstable coupled-bunch modes the most significant term in the summation of Eq.(\ref{alpha}) comes when $q=-1$.

\begin{figure}[!h]
\includegraphics[width=\columnwidth]{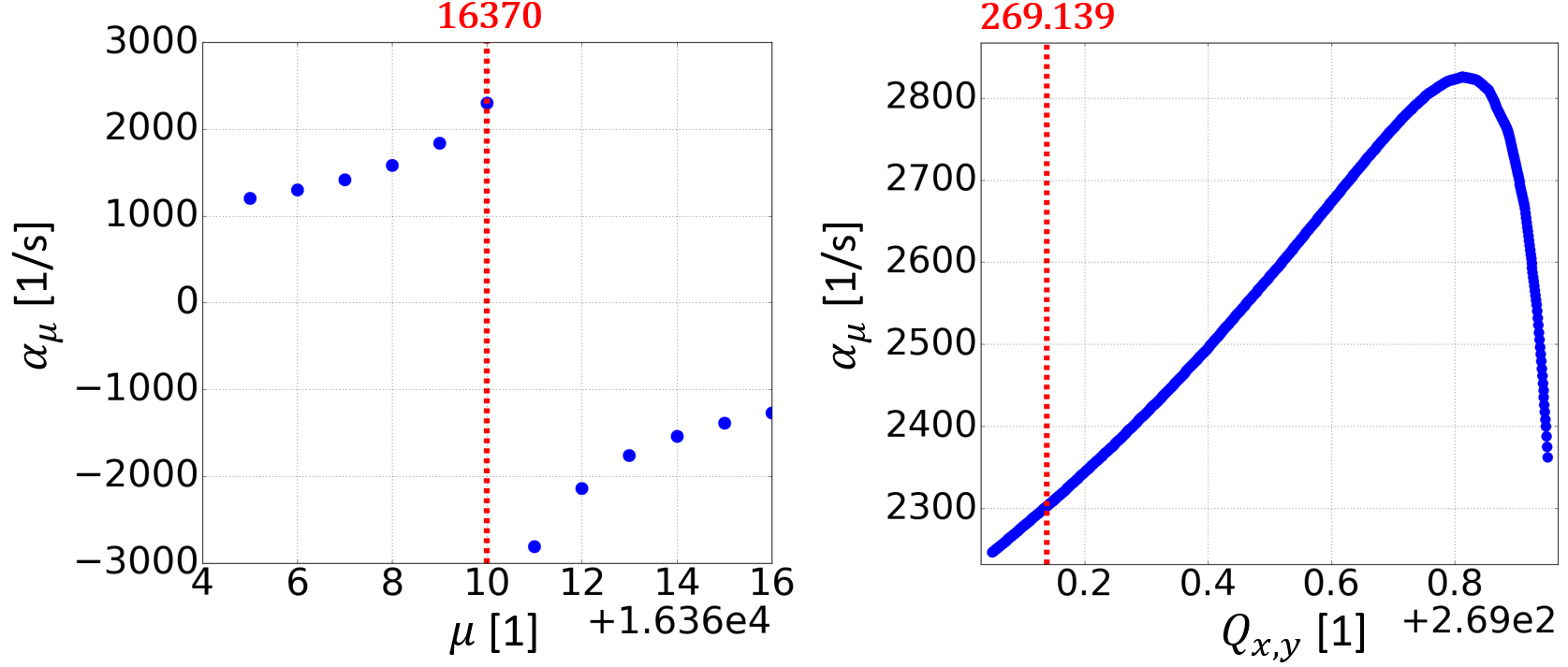}
\caption{Left: TCBI growth rate, as a function of the coupled-bunch mode, obtained with Eq.(\ref{alpha}) assuming $\delta_\text{l}= 1\upmu$m. The most unstable mode is marked by a red line. Right: TCBI growth rate as a function of $Q_{x,y}$ with $\left\lfloor  Q_{x,y} \right\rfloor$=269 and $\mu=16370$. The value $Q_{x,y}^{\text{nom}}$  is marked by a red line. All the other needed parameters are taken from Table 1.}
\label{fig:final}
\end{figure}

Figure \ref{fig:final} (left) shows $\alpha_{\mu}$ as a function of $\mu$ and confirms that $\bar{\mu}$ leads to the largest growth-rate $\alpha_{\bar{\mu}}=2320$ 1/s. 

Figure \ref{fig:final} (right) shows $\alpha_{\mu}$ as a function of $Q_{x,y}$ when $\mu=16370$ and the integer part of the tune is $\left\lfloor  Q_{x,y} \right\rfloor$=269. Note that $\mu=16370$ is the most unstable mode for all these values of $Q_{x,y}$. The plot indicates that the maximum growth rate of 2830 is achieved when $Q_{x,y}=269.812$.

The value $\alpha_{\bar{\mu}}=2320$ calculated for the nominal tune corresponds to a TCBI rise-time of 0.435 ms or 1.33 revolution turns. If $Q_{x,y}=269.812$, then the rise-time is 1.08 turns.

These values for the TCBI rise-time are lower than the SR transverse damping-time by several orders of magnitude. Therefore SR cannot help suppressing this instability. 

Transverse bunch-by-bunch feedback systems are usually used in other lepton factories to counteract TCBI. However these systems cannot act on the short time of one revolution turn. Therefore new challenging feedbacks are required and some schemes have already been proposed \cite{Drago:2017dwa}.

\section{CONCLUSION}

The present contribution showed that the first-designed parameters for the FCC-ee booster cannot provide
stable beams to the main ring. This is due to the resistive-wall impedance which leads to
microwave instability for nominal-intensity beams even if a copper layer is added to the stainless-steel beam
pipe for impedance reduction. Therefore, a second mitigation technique was also taken into account, i.e. the increase of the power lost by
the beam for synchrotron radiation. This second mitigation is in agreement with the current booster baseline plans, which foresee the installation of several wigglers in the machine. Using a proper combination of
parameters, microwave and transverse-mode-coupling instabilities were not observed for nominal-intensity beams. However analytical estimations indicated that the
transverse-coupled-bunch-instability rise-time is only about one revolution turn making necessary the design of new challenging feedback systems.

\section{ACKNOWLEDGEMENTS}
Special thanks go to E. Belli, R. Kersevan, K. Oide and F. Zimmermann for their precious support. This work was partly supported by the European Commission under HORIZON 2020 Integrating Activity project ARIES, Grant agreement $\text{n}\textsuperscript{\underline{o}}$ 730871, and by INFN National committee through the ARYA project.

\bibliographystyle{ieeetr} 
\bibliography{bib} 

\begin{thebibliography}{10}

\bibitem{Abada:2019zxq}
A.~Abada {\em et~al.}, ``{FCC-ee: The Lepton Collider},'' {\em Eur. Phys. J.
  ST}, vol.~228, no.~2, pp.~261--623, 2019.

\bibitem{Harer:2649814}
B.~H{\"a}rer, B.~Holzer, Y.~Papaphilippou, and T.~Tydecks, ``{Status of the
  FCC-ee Top-Up Booster Synchrotron},'' no.~CERN-ACC-2018-111, p.~MOPMF059. 3
  p, 2018.

\bibitem{mauro}
M.~Migliorati {\em et~al.}, ``Collective effects in the booster synchrotron,''
  presented at the FCC Week 2019, Brussels, Belgium, 2019.

\bibitem{Drago:2017dwa}
A.~Drago, ``{Feedback systems for FCC-ee},'' in {\em {Proceedings, 58th ICFA
  Advanced Beam Dynamics Workshop on High Luminosity Circular {$e^+ e^-$}
  Colliders (eeFACT2016): Daresbury, United Kingdom, October 24-27, 2016}},
  p.~TUT3AH9, 2017.

\bibitem{iw2d}
``{CERN IW2D code}.''
  \url{https://twiki.cern.ch/twiki/bin/view/ABPComputing/ImpedanceWake2D}.

\bibitem{Henry:1991xi}
O.~Henry and O.~Napoly, ``{The Resistive pipe wake potentials for short
  bunches},'' {\em Part. Accel.}, vol.~35, pp.~235--248, 1991.

\bibitem{Shaposhnikova:381423}
E.~Shaposhnikova, ``{Signatures of microwaves instability},'' Tech. Rep.
  CERN-SL-99-008-HRF, Feb 1999.

\bibitem{blond}
``{CERN BLonD code}.'' \url{https://blond.web.cern.ch}.

\bibitem{Boussard:872559}
D.~Boussard, ``{Observation of microwave longitudinal instabilities in the
  CPS},'' Tech. Rep. CERN-LabII-RF-INT-75-2. LabII-RF-INT-75-2, CERN, Geneva,
  Apr 1975.

\bibitem{4327284}
R.~H. {Helm}, M.~J. {Lee}, P.~L. {Morton}, and M.~{Sands}, ``Evaluation of
  synchrotron radiation integrals,'' {\em IEEE Transactions on Nuclear
  Science}, vol.~20, pp.~900--901, June 1973.

\bibitem{Sacherer:322645}
F.~J. Sacherer, ``{Transverse Bunched Beam Instabilities - Theory},'' {\em
  Conf. Proc.}, vol.~C740502, pp.~347--351, 1975.

\bibitem{delphi}
``{CERN DELPHI code}.''
  \url{https://twiki.cern.ch/twiki/bin/view/ABPComputing/DELPHI}.

\bibitem{doi:10.1142/5835}
K.~Y. Ng, {\em Physics of Intensity Dependent Beam Instabilities}.
\newblock World Scientific, 2005.

\end{thebibliography}

\end{document}